\documentclass{ws-procs11x85}

\usepackage{balance}

\newcommand{\smallw}{{\scriptscriptstyle W}} %
\newcommand{\smallh}{{\scriptscriptstyle H}} 
\newcommand{\smallp}{{\scriptscriptstyle P}} %
\newcommand{\mh}{M_\smallh}
\newcommand{\mw}{M_\smallw}

\newcommand\gsim{\mathop{\mbox{\vbox{\hbox{$>$} \vskip -9pt \hbox{$\sim$}
             \vskip -3pt  }}}}
\newcommand\lsim{\mathop{\mbox{\vbox{\hbox{$<$} \vskip -9pt \hbox{$\sim$}
             \vskip -3pt  }}}}
\newcommand{\beq}{\begin{equation}}                
\newcommand{\eeq}{\end{equation}}        
\newcommand{\bea}{\begin{eqnarray}}               
\newcommand{\eea}{\end{eqnarray}}        
\makeindex
\begin{document}

\title{THE TOP PRIORITY:
PRECISION ELECTROWEAK \\  PHYSICS FROM LOW TO  HIGH ENERGY}

\author{P.~GAMBINO}

\address{Theory Division - CERN,
CH 1211, Geneve 23, Switzerland\\
E-mail: paolo.gambino@cern.ch\\
(Present address: INFN, sez.\ di Torino,
    v.~Giuria 1, 10125 Torino, Italy)
}

\twocolumn[\maketitle\abstract{
Overall, the Standard Model describes  electroweak precision data rather well. 
There are however a few areas of tension (charged current
universality, NuTeV, $(g-2)_\mu$, $b$ quark asymmetries), 
which I review 
emphasizing recent theoretical and experimental progress. I also
discuss what precision data tell us about the Higgs boson and new
physics scenarios. In this context, 
the role of a precise measurement of the top mass is crucial.
}]

\baselineskip=13.07pt

Precision electroweak physics lies at the intersection of 
many specialized fields and involves experiments performed at hugely
different energies. In testing the consistency between all  data 
within  the Standard Model (SM) framework, we hope to uncover signs
of physics beyond the SM. However, as we will see, the main problem is the
precision of the theoretical predictions with which we  confront the 
experimental data. Almost invariably, long-distance hadronic
interactions enter the game, so we often take great pains to try to make
sense of extremely precise experiments.

In the following I will try to summarize the main recent progress in
the field, concentrating on the unsettled questions.  For some of the
topics I will not have space to cover, see the References\rlap.\,\cite{pippa,erler}

\section{Parity Violation in M\o ller Scattering}
Let us start from low-energy experiments. 
The E158 experiment at SLAC\cite{E158} has measured for the first time parity
violation (PV) in polarized M\o ller ($e^-e^-$) scattering. The PV
asymmetry $A_{LR}= (\sigma_L-\sigma_R)/ (\sigma_L+\sigma_R)$ is
extremely small in the SM, $\approx 10^{-7}$, 
due to an extra suppression factor $1/4-\sin^2\theta_W$. It can be
measured at SLAC  thanks to the huge luminosity and the high
polarization of the beam. $A_{LR}$ is very sensitive to
$\sin^2\theta_W$ and the goal of E158 is to measure it with  8\%
precision, equivalent to an error of 0.001 on $\sin^2\theta_W$.
Such a precision is not competitive with LEP and SLD determinations, but
one should keep in mind that a low-energy  measurement would test completely
different radiative corrections, and would be sensitive to new physics
complementary or orthogonal to collider experiments.

E158 is currently performing a last and third run and expect to
be able to reach the aimed precision.
The preliminary result of Run I (at $Q^2=0.027$~GeV$^2$),
\[
A_{LR}=[151.9\pm 29.0(stat) \pm 32.5(syst)] \times
10^{-9},
\]
translates into $\sin^2\hat\theta_W^{\overline{\rm MS}}(M_Z)=0.2296\pm
0.0038$, in good agreement with the global average, $\sin^2\hat
\theta_W^{\overline{\rm MS}}(M_Z)=0.2312\pm 0.0003$. Radiative
corrections\cite{Moeller} 
reduce $A_{LR}$ by about 40\%. A  large theoretical uncertainty
comes from the $\gamma-Z$ hadronic vacuum polarization, which cannot
be computed perturbatively. The current estimate, inducing $\approx 5\%$
error  on $A_{LR}$, can and should be updated in view of E158's final result,
expected next year. 

\section{Universality of Charged Currents}
This is a very old subject\rlap.\,\cite{sirlin} Universality 
in the leptonic sector is verified at the 0.2\% level\rlap.\,\cite{marciano} 
Charged currents in the quark sector, on the other hand, involve also
the CKM matrix elements. One can however test accurately
the unitarity relation 
\beq\label{unitarity}
|V_{ud}|^2 + |V_{us}|^2 + |V_{ub}|^2=1.
\eeq
Since the last term on the lhs is $O(10^{-5})$, 
the test concerns the consistency of
Cabibbo angle measurements from $V_{ud}$ and $V_{us}$. 

The most precise method to measure $V_{ud}$ is to use Superallowed
Fermi Transitions, i.e. $0^-\to 0^-$ nuclear $\beta$
decays. There are  several experiments in good agreement, yielding 
$\delta V_{ud}\sim 0.0005$. Neutron $\beta$ decay is also becoming
competitive: the present $\delta V_{ud}\sim 0.0013$ will be improved at
PERKEO\cite{perkeo}. A promising mode is pion decay, currently at 
$\delta V_{ud}\sim 0.005$, which is theoretically cleaner and will
soon be improved by PIBETA\cite{pibeta}. 
The consistent picture that emerges from
these experiments can be expressed, using Eq.~(\ref{unitarity}), as
\beq\label{vus_un}
|V_{us}| ({\rm unitarity})=0.2269\pm 0.0021.
\eeq

The most precise direct measurement of $|V_{us}|$ is given on the
other hand by 
$K\to\pi\ell\nu$ decays ($K_{\ell 3}$). 
Here the experimental situation is not as consistent as for $V_{ud}$:
the recent E865 result for $K^+$ decays disagrees 
with  a series of old experiments by more than 2$\sigma$. While the E865 result
agrees well with the unitarity prediction, Eq.~(\ref{vus_un}), the
older results and a recent preliminary $K^0$ measurement by KLOE
all yield a smaller Cabibbo angle. Upcoming
analyses from KLOE, NA48, and KTeV should tell us whether
grossly underestimated isospin breaking corrections are the cause of
this situation, or there is an experimental problem. 
Averaging the old published data only, one obtains
\[
|V_{us}|_{K_{\ell 3}}=0.2201\pm 0.0024,
\]
but the result changes little if one includes also E865 and KLOE results.
Alternative promising strategies to extract $|V_{us}|$ are provided by $\tau$
and hyperon decays. In particular, measurements of the $\tau$ spectral
functions at the $B$ factories will make the first method competitive
with $K_{\ell 3}$, while the use of hyperon decays 
requires a careful assessment
of SU(3) breaking effects, which could be helped by lattice simulations.

In summary, a puzzling violation of unitarity persists at the level of
$\sim 2\sigma$, despite new data. Fortunately, upcoming experimental
results are likely to shed light on this problem.
For a more detailed discussion, see elsewhere\rlap.\,\cite{ckm}

\section{The NuTeV Electroweak Result}
 NuTeV measures ratios of Neutral  (NC) to Charged Current (CC)
 cross sections in $\nu N$ DIS\rlap.\,\cite{nutev} 
Ideally, in the parton model with only one generation of quarks
and an isoscalar target
\begin{eqnarray}
&&R_\nu \equiv \frac{\sigma(\nu { N}\to \nu X)}{\sigma(\nu { N}\to \mu X)} =
g_L^2 + { r} g_R^2, \nonumber\\
&&R_{\bar{\nu}} \equiv \frac{\sigma(\bar\nu { N}\to \bar\nu
X)}{\sigma(\bar\nu { N}\to \bar\mu X)} =
 g_L^2 + \frac{1}{ r} g_R^2,\label{rdef}
\nonumber
\end{eqnarray}
where
$ r \equiv  \frac{\sigma(\bar{\nu}{ N}\to \bar\mu
X)}{\sigma({\nu}{ N}\to \mu X)}$ and $g_{L,R}^2$ are average
effective left- and right-handed $\nu$-quark couplings. 
The actual experimental  ratios
$R_{\nu,\bar{\nu}}^{exp}$ differ from $R_{\nu,\bar\nu}$ because
of $\nu_e$  contamination, experimental cuts, NC/CC misidentification,
the presence of second generation
quarks, the non-isoscalarity of steel  target, QCD and electroweak
corrections, etc.. In the NuTeV analysis, a
Monte Carlo including most of these effects 
relates $R_{\nu,\bar{\nu}}^{exp}$ to $R_{\nu,\bar{\nu}}$. 
It is useful to note that 
most uncertainties and $O(\alpha_s)$ effects drop out in the
Paschos-Wolfenstein (PW) ratio\cite{PW}
\[\label{eq:PW}
R_{\smallp\smallw}\! \equiv\!\frac{R_\nu - { r} R_{\bar{\nu}}}{1- r}\! =\!
\frac{\sigma(\nu { N}\to \nu X)-\sigma(\bar\nu { N}\to
\bar\nu X)}{\sigma(\nu { N}\to \ell X) - 
\sigma(\bar{\nu}{ N}\to \bar{\ell}X)}
\]
which equals ${ g_L^2- g_R^2} %
= \frac{1}{2}-\sin^2 \theta_{\rm W}$ and therefore could  provide a clean
measurement of $\sin^2 \theta_{\rm W}$, if experimentally accessible.
NuTeV do not measure $R_{\smallp\smallw}$ directly,
but, using the fact that
$R_{\bar{\nu}}$ is almost insensitive to $\sin^2 \theta_{\rm
W}$, they  extract from it 
the main hadronic uncertainty, an effective charm mass. 
The weak mixing angle 
is then obtained from $R_{\nu}$.
In practice, NuTeV fit for $m_c^{\rm eff}$ and $\sin^2\theta_{\rm W}$.
 This procedure certainly approximates a measurement of
 $R_{\smallp\smallw}$, but it is not clear to what extent exactly.

The NuTeV result provides a test of the on-shell 
$s_W^2\equiv 1- M_W^2/M_Z^2$ definition of $\sin^2\theta_W$:
\begin{equation}\label{result}
s_W^2({\rm NuTeV})=0.2276\pm0.0013\pm0.0006\pm0.0006,
\end{equation}
where the three errors are 
statistical, systematic and theoretical respectively.
Because of accidental cancellations, the choice of the on-shell scheme 
implies very small top and Higgs mass dependences in Eq.~(\ref{result}).
The above value must be compared to the result of the global fit, 
 $s_W^2=0.2229\pm 0.0004$, which is 2.8$\sigma$ away. 

NuTeV works at Leading Order (LO) in QCD in 
the context of a {\it cross  section model} which effectively introduces
some Next-to-Leading-Order (NLO) improvement. They use
LO PDF's self-consistently fitted in the experiment, with little 
external input. There are a number of theoretical systematics which could have
been underestimated in Eq.~(\ref{result}), and considerable work has been
devoted to study the most obvious among them.
\begin{itemize}
\item [{\it i)}] {\it
  Uncertainties in the parton distribution functions}
 (PDF's): 
neglecting for the moment asymmetric sea contributions (see later)
they are small in $R_{\smallp\smallw}$ with the cuts used.\cite{noi,NLOQCD2}

\item [{\it ii)}] {\it NLO QCD corrections}:\cite{NLOQCD2}$^-$\cite{NLOQCD3}
vanish in $R_{\smallp\smallw}$, 
and effects introduced by asymmetric cuts and
differences in the $\nu,\bar\nu$ energy spectra seem small. 
Again,  this refers to the ideal observable $R_{PW}$.
Only a complete NLO analysis can ensure that the same conclusions
apply to the NuTeV fit. For instance,  
the phenomenological
cross section model used by NuTeV may distort in an important way
cancellations among QCD corrections\rlap.\,\cite{NLOQCD3} 
Estimating the actual effect on
$s_W^2$ would require refitting the PDF's at NLO.
In summary, the analysis needs to be consistently upgraded to NLO, and 
the NuTeV collaboration is investigating this possibility.
\begin{figure}[t]
\center{\psfig{figure=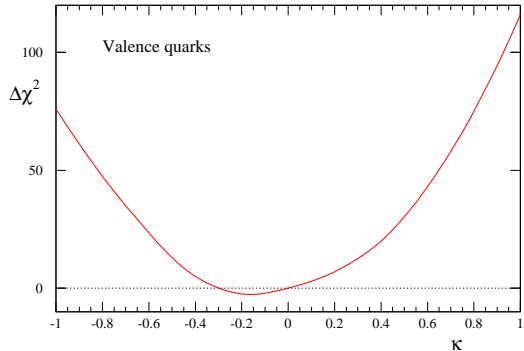,width=3.truein}}
\caption{MRST fit of  isospin violation in valence PDF's.
\label{fig:mrst} \vspace*{12pt}}
\end{figure}

\item [$iii)$]{\it  Electroweak  corrections} (mainly photonic):
the NuTeV analysis is largely based on very old code, 
which needs to be checked against recently developed
tools\rlap.\,\cite{ew} 

\end{itemize}

{\subsection{ Asymmetric Sea}}
I have so far  used the assumptions, 
generally made in the extraction of PDF's from the data, 
of isospin symmetry  and of a symmetric strange and charm sea ($s=\bar
s$, $c=\bar c$). 
If we drop these assumptions, the PW relation is explicitly violated by new
 terms\cite{noi}
\begin{equation}
R_{\smallp\smallw}= 
\frac12 -s_W^2 + \frac{\tilde{g}^2}{Q^-} \,( u^- - d^- +c^- - s^-),\label{viol}
\end{equation}
where $q^-$ is 
the asymmetry in the momentum carried by the quark species $q$ 
in an isoscalar target, $q^-=\int_0^1  x \,[q(x)- \bar{q}(x)] \,dx$,
$\tilde{g}^2\approx 0.23$ a coupling factor, 
and $Q^-=(u^-+d^-)/2\approx 0.18$.
The non-isoscalarity of the target gives a contribution
to $u^- - d^-$ that is obviously taken into account by
NuTeV, although the uncertainty on this correction seems to have been
somewhat underestimated\rlap.\,\cite{kulagin} There are, however, less standard
and potentially more dangerous  contributions:
there is no reason in QCD to expect $s^-\!=\! 0$, 
and  for an isoscalar target $u^-\!-\!d^-$ is of the order of
the isospin violation. Eq.~(\ref{viol}) tells us that 
even quite small values of these two
asymmetries could change significantly  the value of $s_W^2$
measured by NuTeV\rlap.\,\footnote{ 
These effects are somewhat diluted in the actual
NuTeV analysis compared to the direct use of Eq.~(\ref{viol})\rlap,\,\cite{nutev2} 
precisely because NuTeV differs from a 
measurement of $R_{\smallp\smallw}$.}

A violation of isospin of the form $u_p(x)\neq d_n(x)$ would
induce a $u^-$ different from $d^-$ even in an isoscalar target and
affect the PW relation according to Eq.~(\ref{viol}). A rough estimate
for its size is 
$(m_u-m_d)/{\Lambda_{QCD}}\approx  1\%$.
This  could explain a fraction of the 
anomaly -- about a third, according to Eq.~(\ref{viol}).
Isospin  violation  is very weakly constrained by 
experiment, as demonstrated by a new MRST analysis\rlap.\,\cite{thorne}
MRST have performed a global fit to the PDF's deforming the valence
distributions by a contribution proportional to a function, $f(x)$,
with zero first moment: $u^-_n(x)= d_p^-(x) + \kappa f(x)$ and
 $d_n^-(x)= u_p^-(x) - \kappa f(x)$. The fit to $\kappa$, shown in
Fig.~1, gives a mild indication for a negative $\kappa$, but with very
large uncertainty (MRST use $\Delta\chi^2=50$ to define a 90\% CL).
 The central value $\kappa\approx -0.2$ corresponds
to a reduction of the NuTeV anomaly by about a third, and has the
expected order of magnitude.   
Amusingly, the MRST central value leads to a shift in $s_W^2$ 
very close   to that of a recent analysis 
in the context of nucleon models\rlap.\,\cite{londergan} 
Using similar models, NuTeV claim a much smaller isospin breaking 
shift\rlap.\,\cite{nutev2}
In any case, it is clear that model calculations\rlap,\,\cite{isospin}
though sometimes useful to understand the size of an effect, cannot be relied
upon for a precision measurement. We are therefore left with a
substantial uncertainty unaccounted for in Eq.~(\ref{result}).
\begin{figure}[t]
\ \ \psfig{figure=figs/NumAsym1.eps,width=2.8truein}\\
\psfig{figure=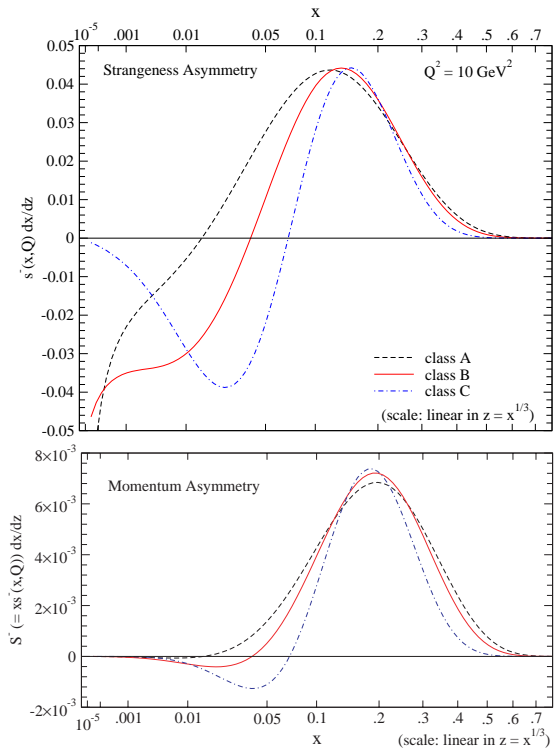,width=2.9truein} 
\caption{CTEQ fit of the strangeness asymmetry using different low-$x$
  behaviors. 
\label{fig:cteq1} }
\end{figure}

What do we know about the strange quark asymmetry?
An asymmetry $s^-$ of  the sign needed to explain NuTeV 
can be induced non-perturbatively
({\it intrinsic strange}) by fluctuations of the kind
 $p\!\!\leftrightarrow\!\!\! \Lambda\, K^+$\rlap.\,\cite{brodsky} 
Unfortunately, the strange quark sea is
mainly constrained by (mostly old) $\nu N$ DIS data, which are
 usually not included in standard PDF's fits. In fact,  
MRST and CTEQ use an {\it ansatz}
$s\!=\!\bar{s}\!=\!(\bar{u}+\bar{d})/{4}$.
Barone {\em et al.} (BPZ)\cite{BPZ} reanalyzed, a few years ago,
a host of  $\nu N$ DIS 
together with  $\ell N$ and Drell-Yan data at NLO. 
Allowing for a  strange asymmetry improved the BPZ best fit drastically
and could explain a large fraction  of the NuTeV discrepancy. 
The result,  { $s^-\approx0.0018\pm 0.0005$},
was compatible with theory estimates\cite{brodsky} and was driven by
cross section measurements by CDHSW ($\nu$N) and BCDMS ($\mu\,p$). 
\begin{figure}[t]
\center{\psfig{figure=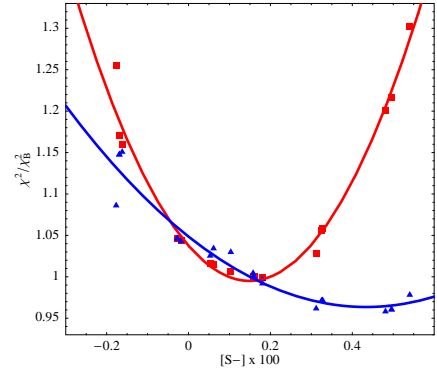,width=2.2truein}}
\caption{CTEQ fit of the strange momentum asymmetry $s^-$.}
\vspace{-0.5cm}
\label{fig:cteq} 
\end{figure}
The BPZ analysis was recently updated with the inclusion of CCFR cross
sections, leading to a quite different result,  $s^-\approx0.0002\pm 0.0004$.

The inclusive analysis pioneered by BPZ should however be supplemented 
by data on  dimuon events (tagged charm production),  a rather sensitive
probe of the strange sea. The most precise dimuon data come from the
CCFR/NuTeV Collaboration\rlap,\,\cite{nutev_dimuon} which has 
analyzed them at LO with the specific aim of constraining the strange
asymmetry. Their result, $s^- =-0.0027\pm 0.0013$, 
would {\it increase} the anomaly to 3.7$\sigma\rlap,\,$\cite{nutev2} 
but it suffers from
various shortcomings, detailed in the note added 
to S.~Davidson {\em et al.}\cite{noi} and elsewhere\rlap.\,\cite{cteq}
The main problem is in the parameterization, which does not satisfy
the condition
\beq\label{s_zero}
\int_0^1 dx [s(x)-\bar s(x)]=0
\eeq
that ensures zero  strangeness quantum number for the nucleon.
As the dimuon data are concentrated at $x<0.3$, the evidence for a
small negative strange asymmetry at low $x$ would imply, if 
the condition given by Eq.~(\ref{s_zero}) is imposed,  a {\it positive}
asymmetry at large $x$, and hence a {\it positive} momentum asymmetry.
This is illustrated in Fig.~2\rlap,\,\cite{cteq} which shows strange
asymmetries with the above qualitative features but different shapes. 
The NuTeV analysis of dimuon data is not reliable.

A dedicated global fit that employed both inclusive and 
dimuon data in a consistent way
was therefore necessary. The CTEQ Collaboration has presented at this
conference the preliminary results of one such analysis\rlap.\,\cite{cteq}
The inclusion of the CCFR-NuTeV dimuon data in the CTEQ global fit 
is presently done using NuTeV software developed at  LO in QCD. 
Dimuon data  are therefore included at LO, 
which should not influence the main qualitative conclusions.
CTEQ explored the full range of parameterizations of $s(x)-\bar s(x)$ 
that satisfy Eq.~(\ref{s_zero}), studying for instance different low-$x$ 
behavior, as shown in Fig.~2. They perform a new global fit to all
PDF's using all available inclusive and dimuon 
data, although they do not reanalyze old $\nu N$ data in detail, as was done by BPZ.
The preliminary 
result of the $s^-$ fit is shown in Fig.~3 for the best performing
(class B) parameterization. While inclusive data alone show only a mild
preference for a positive $s^-$, the dimuon data have real discriminating
power. The central value of the global class B fit is $s^-\approx
0.002$, and corresponds to the indicated line in Fig.~2.
In general, all acceptable fits have central values
$0.001<s^-<0.003$. Negative $s^-$ are disfavored, but $s^-=0$
cannot be excluded. CTEQ estimate that 
the likely impact on the NuTeV $s_W^2$ extraction would be a reduction
of $s_W^2$ by 0.0012 to 0.0037.
Note that if a strange asymmetry shifted $s_W^2$ by $0.002\pm 0.002$, the
NuTeV result would be at 1$\sigma$ from the SM.
Although a more detailed study 
is under way with the active participation of the NuTeV
Collaboration, two firm conclusions are that: $i)$ the strange
asymmetry is a strong candidate to
explain  part or most of the NuTeV anomaly; and $ii)$ 
one cannot avoid the related,  substantial uncertainty.

Given the present understanding of hadron structure, $R_{\smallp\smallw}$
does not seem to be a good place for high
precision electroweak physics. In fact, 
the relevant momentum asymmetries in the quark sea induce an error in
the extraction of  $s_W^2$ of the same order as the experimental error.
Improved analyses of dimuon data would certainly constrain $s^-$
better, and  data from CHORUS might also be useful -- if not for
measuring $s^2_W$, at least for constraining the sea asymmetries\rlap.\,\cite{chorus}
Useful input could also come from associated charm-$W$ production at
the Tevatron and RHIC. In the long term, a
 precise $s(x), \bar{ s}(x)$ determination will be possible at a 
neutrino factory\rlap.\,\cite{nufact}

I should also mention that several attempts at explaining the NuTeV
anomaly with nuclear effects like nuclear shadowing
have been made\rlap,\,\cite{nuclear} but 
no convincing case has so far  been presented.

\vspace{0.5cm}
\subsection{New Physics vs NuTeV}

A new physics explanation of the NuTeV anomaly requires a $\sim 1$-2\% effect, 
and naturally calls for  tree level physics. It is very difficult to build
realistic models that satisfy all present experimental  constraints
and explain a large fraction of the anomaly\rlap.\,\cite{noi}

In particular, Supersymmetry, with or without R-parity, cannot help,
because it is strongly constrained by other precision
measurements (often at the 
$10^{-3}$ level) and by direct searches\rlap.\,\cite{noi,ramsey}
The same is generally true of models inducing only oblique corrections
or only anomalous $Z$ couplings\rlap.\,\cite{noi} Realistic and well-motivated 
examples of the latter are 
models with $\nu_R$ mixing\rlap.\,\cite{noi,babu}
Models with $\nu_R$ mixing {\it and} oblique corrections have been
considered by W.~Loinaz {\em et al.}\cite{loinaz} and found to fit well all
data including NuTeV\rlap.\,\footnote{Can the 
necessary oblique corrections be provided by a heavy SM Higgs boson?
No, the only way  to obtain an acceptable fit with a preference for 
both $\nu$ mixing and  a heavy Higgs is to exclude $M_W$ from the 
data\rlap.\,\cite{loinaz}
However, 
solving the NuTeV anomaly at the expense of the very precise measurement of 
$M_W$ is hardly an improvement.} 
However finding sensible new physics that provides  
oblique corrections in the preferred range is far from obvious.

On the other hand, the required new physics can be parameterized  by
a contact interaction of the form
$[\bar{L}_2 \gamma_\mu L_2][\bar{Q}_1
\gamma_\mu Q_1]$. This operator might be induced by different kinds of
short-distance physics. Leptoquarks generally also induce another
operator which over-contributes to $\pi\to \mu \bar\nu_\mu$, or have
the wrong sign, but SU(2) triplet leptoquarks with non-degenerate
masses could fit NuTeV, albeit not very naturally.
Another possible new physics contribution inducing
 the above contact interactions is
an unmixed $Z'$ boson. It could be either light ($2\lsim M_{Z'} \lsim
10$~GeV) and super-weakly coupled, or heavy ($M_{Z'} \gsim 600$~GeV). 
The $Z'$ must have very small mixing with the $Z^0$ because of the
bounds on oblique parameters and on the anomalous $Z$ 
couplings\cite{noi,erlerZ} (see  E.~Ma and D.~P.~Roy\cite{ma} 
for an explicit $L_\mu-L_\tau$ model and
R.~S.~Chivukula and E.~H.~Simmons\cite{chivu} for technicolor models).

\section{The Ups and Downs of {\boldmath $(g-2)_\mu$}}
The anomalous magnetic moment of the muon is an excellent place to
look for new physics: it probes unexplored loop effects proportional to
 $m_\mu^2/\Lambda^2$, where $\Lambda$ is the mass scale characteristic 
of the new physics. Given the present experimental resolution, 
in order for us to observe large deviations from the SM, the 
new physics we need must have a chiral
enhancement, of the kind naturally emerging in Supersymmetric models
with large $\tan\beta$. Conversely, no deviation from the SM would impose 
severe constraints on these models. This is at the origin of the great 
attention this observable has recently received.
\begin{figure}[t]
\center{\psfig{figure=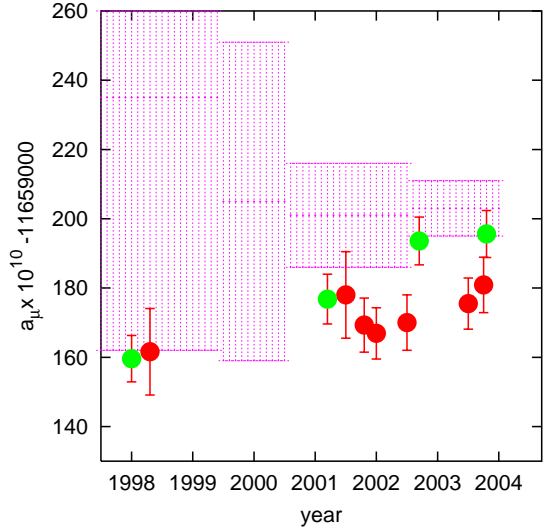,width=3.truein}}
\caption{Historical evolution of the measurement and prediction of
  $a_\mu$. The lighter (green) dots represent estimates based also on $\tau$
  decay data.  The compilation of theoretical estimates is not complete. }
\vspace{-0.1cm}
\label{fig:g-2hist} 
\end{figure}

The last few years have seen a dramatic  progress in 
the measurement of $a_\mu$, driven by the $g-2$ experiment at Brookhaven.
The present world average
\[
a_\mu({\rm w.a.} ) =11659203(8) \times 10^{-10}
\]
is dominated by their latest $\mu^+$ result\cite{anomuon}, released in 2002. The 
results of the 2001 Run, performed with $\mu^-$,  should reduce the
error by $\sim 30\%$ and are expected soon.
\begin{figure}[ht]
\center{
\psfig{figure=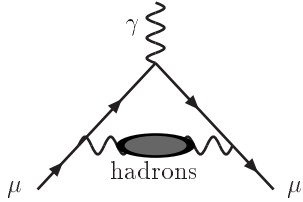,width=1.6 truein}}
\caption{Lowest order vacuum polarization insertion.}
\vspace{-0.3cm}
\label{fig:diags} 
\end{figure}

Figure~4 summarizes the evolution of the measurement and of the theoretical
estimates of $a_\mu$. As you will see in a moment,
 the theoretical prediction of this quantity depends heavily on 
other experimental results, so the ups and downs are mostly due to the
evolution of data and to the corrections of some unfortunate mistakes. 

While most of us have computed the lowest-order  QED contribution to 
$a_\mu$ at  graduate school, a calculation of $a_\mu$ at the current  level
of precision is a very involved and sophisticated enterprise (there
are excellent reviews\rlap,\,\cite{g-2reviews} with references to the
original literature). Here I will concentrate only on the general
aspects and on recent developments. The various contributions to
$a_\mu$, listed with their estimated errors, are:
\bea
a_\mu&=& 11\, 658\,470.35(28) \times 10^{-10} \nonumber \, ({\rm QED})\\
&& +694(7) \times 10^{-10} \, ({\rm had, Leading\,Order})\nonumber\\
&& -10.0(6) \times 10^{-10} \, ({\rm had, Higher\,Order})\\
&& +8(4) \times 10^{-10} \, ({\rm had, Light\, by\, Light})\nonumber\\
&& +15.4(2)\times 10^{-10} \, ({\rm EW})\nonumber
\eea
The main component comes from  QED without hadronic
loops. The four-loop contribution\cite{kinoshita} is not so small,
$\sim 40 \times
10^{-10}$, and has never been checked. But these heroic calculations at
least can be done. Not so for the hadronic contributions:
 hadronic loops enter the second order diagram of
Fig.~\ref{fig:diags} and are characterized by the scale
$\Lambda_{QCD}\approx 300$~MeV. They provide the largest uncertainty
to the determination of $a_\mu$. As the energy scale is too low to employ
perturbative methods, the usual route is to use a dispersive integral
of the vector spectral function, 
\begin{figure}[t]
\center{
\psfig{figure=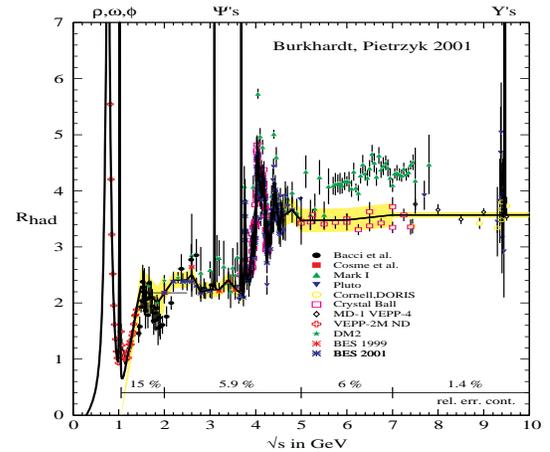,height=6cm,width=7cm}}
\caption{The spectral function.
\label{fig:spectral} \vspace*{12pt}}
\end{figure}
\beq\label{int}
a_\mu^{\rm LO,had}= \frac1{4\pi^3} \int_{4m_\pi^2}^{\infty}
R_{\rm had}(s) K(s) ds
\eeq
where  the spectral function $R_{had}(s)$ is measured
from the total hadronic cross section in $e^+e^-$ collisions.
A number of experiments have contributed to its measurement, most
recently CMD-2, SND, and BES, leading to the situation summarized in
Fig.~\ref{fig:spectral}.  
Different strategies are also available to combine the data and their
errors -- see the References\cite{HMNT}$^-$\cite{jeg} for the most recent and complete analyses.
Because of the weight function $K(s)$, the integral given by Eq.~(\ref{int}) 
is dominated by the low energy region, and in particular
by the $\rho$ resonance in the  $\pi\pi$ channel. Indeed, the pion
form factor (see Fig.~\ref{fig:rho}) 
alone contributes more than 70\% of $a_\mu^{\rm LO,had}$. The 
recent CMD-2 reanalysis\cite{CMD2} of their very precise $\pi\pi$ data, 
with a revised treatment of QED corrections, is
therefore of the utmost importance. It is included in the following
\begin{figure}[t]
\center{
\psfig{figure=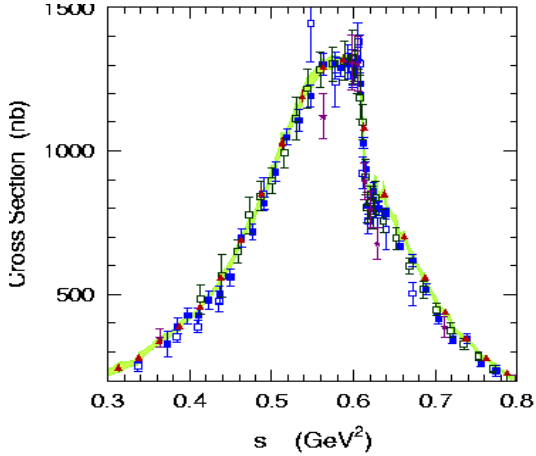,height=6cm,width=7cm}}
\caption{The pion form factor.
\label{fig:rho} \vspace*{12pt}}
\end{figure}
updated estimates:\cite{teubner}$^-$\cite{jeg}
\bea
a_\mu^{\rm LO,had}({\rm HMNT})&=&(691.8\pm 5.8_{exp}\pm2.0_{r.c.})\times 10^{-10} \nonumber\\
a_\mu^{\rm LO,had}(\rm DEHZ)&=&(696.3\pm 6.2_{exp}\pm3.6_{r.c.})
\times 10^{-10} \nonumber\\
a_\mu^{\rm LO,had}(\rm GJ)&=&(694.8\pm 8.6)\times 10^{-10}
\eea
where the r.c.\ error is mostly due to uncertainty in 
correcting old data for missing radiative corrections. Adding
all other SM contributions, this translates into a 1.9-2.5$\sigma$
discrepancy between SM prediction and experiment.

\begin{figure}[h]
\begin{center}
\psfig{figure=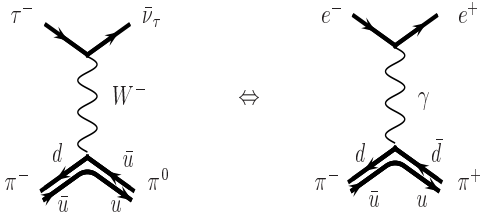,height=3.cm,width=6.4cm}
\end{center}
\caption{The diagrams relating  $\tau$ hadronic decays to the $e^+e^-$ hadronic
cross section.
\label{fig:cvc} }
\end{figure}

A second  way of measuring the spectral function in the crucial region
 below 1.8~GeV consists of 
relating the $\tau$ hadronic decays to the $e^+e^-$ hadronic
cross section using CVC and isospin symmetry, as schematically
illustrated in Fig.~\ref{fig:cvc}.
This method has been explored by M.~Davier\cite{Davier}
 {\it et al.\ } with data from Aleph,
CLEO, and Opal. A series of
corrections have been implemented\cite{cirigliano}, leading to 
\[
a_\mu^{\rm LO,had}(\rm DEHZ,\tau)=(709.0\pm 5.1_{exp}\pm1.2_{r.c.}\pm 2.8_{SU(2)})
\]
where the last uncertainty refers to the isospin corrections. This
determination is competitive with $e^+e^-$ and leads to a prediction
of $a_\mu$ in much better agreement with experiment (0.7$\sigma$).
\begin{figure}[t]
\center{
\psfig{figure=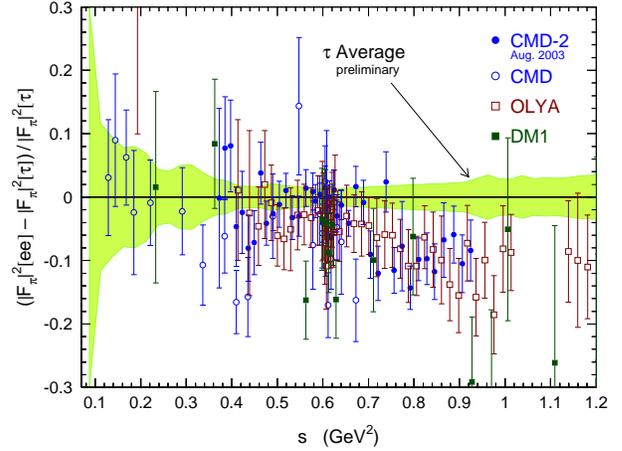,height=6.1cm}
}
\caption{Comparison of the spectral function extracted from $\tau$
  decay and $e^+e^-$ data (relative difference).}
\vspace{-0.5cm}
\label{fig:tau} 
\end{figure}
Figure~\ref{fig:tau} from M.~Davier\cite{Davier} {\it et al.}
shows a comparison of the spectral function
extracted from $e^+e^-$ and $\tau$ data: although the CMD2 revision  has
much improved the situation below 850~MeV, there is still a
discrepancy between 0.85 and 1~GeV. The problem could be in the data
or in the theoretical framework. While a
 recent paper advocates the second possibility\rlap,\,\cite{jeg,bing} 
hinting at underestimated isospin breaking, an important
 check of the CMD2 $e^+e^-$ results has come from the first results of a third
 method to measure the spectral function, the radiative return.

\begin{figure}[h]
\begin{center}
\psfig{figure=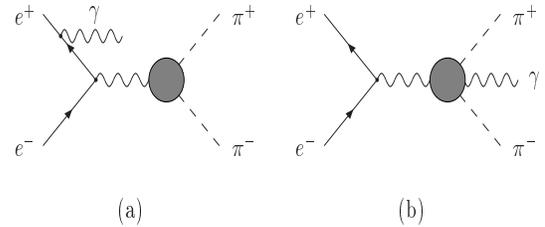,height=3.5cm,width=7.5cm}
\end{center}
\caption{Examples of ISR and FSR.
\label{fig:isr} }
\end{figure}

The idea behind  radiative return is 
that a photon radiated off the initial $e^+$ or $e^-$ (ISR) reduces
the effective energy of the collision, see Fig.~\ref{fig:isr}.
Provided the photon momentum is measured, a fixed energy collider can
investigate a whole $q^2$ range, with obvious advantages over the
energy scan experiments. The large luminosities at DA$\Phi$NE
and at the $B$-factories compensate the radiative suppression. 
The potential pollution from FSR at low-energy
(see Fig.~10(b)) is circumvented by kinematic
cuts. Radiative corrections\cite{phokara}
 play a crucial role here, as they do anyway in
the energy scan case. KLOE
has announced\cite{kloe} the first preliminary results of radiative return: the
contribution of the two pions channel to $a_\mu^{\rm LO,hadr}$
in the range $0.37<q^2< 0.93$~GeV$^2$ is, in units $10^{-10}$,
 $378.4\pm0.8_{stat}\pm
4.5_{syst}\pm 2.6_{th} \pm 3.8_{FSR}$, to be compared with the {\it new} CMD2 result
 $378.6\pm2.6_{stat}\pm 2.2_{syst\& th}$. KLOE agrees well with CMD2.
The systematic error will
soon be further reduced. Radiative return analyses are
also expected from Babar, Belle and CLEO-c.
\begin{figure}[t]
\center{
\psfig{figure=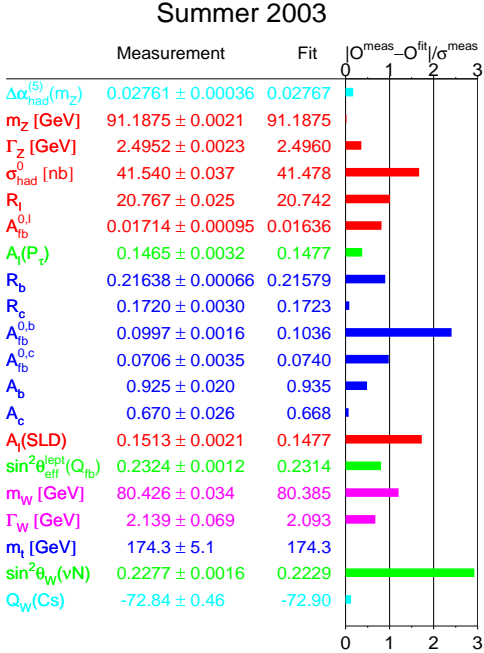,width=2.6truein,height=3.5truein}}
\caption{Pulls in the summer 2003 fit by the LEP Electroweak Working Group.}
\vspace{-0.4cm}
\label{fig:pulls} 
\end{figure}

\section{ The SM Fit and the Higgs Boson Mass}
The latest compilation of electroweak data of the LEP Electroweak
Working Group\cite{pippa} is shown in Fig.~\ref{fig:pulls}, where the
data are compared with the results of a global fit.
The main changes with respect to last year are: a revised (lower) $M_W$ 
value from Aleph, that draws the world average down by $0.5\sigma$, to 
$M_W(w.a.)=80.426\pm 0.034$~GeV and improves the consistency of the
global fit; small shifts in the heavy flavor
observables; and a new value of atomic PV, due to revised (and hopefully
converging) theoretical calculations. The value for $M_t$, 174.3$\pm
5.1$~GeV, is the old one, and does not include the new D0 analysis\rlap.\,\cite{azzi}
Also the value of  $\alpha(M_Z)$, from the conservative 
estimate\rlap,\,\cite{BP} has not yet been updated 
to reflect the new CMD2 data,  a rather small effect anyway
(the new value is $\Delta\alpha_{had}=0.02768\pm 0.00036$).
Indeed, the spectral function discussed in the previous section enters
also the determination of $\alpha(M_Z)$, but higher energy data have
more weight. Considerable progress has been achieved in the
last few years, and this uncertainty is no more a bottleneck for the
present bounds on $M_H$. 
An alternative analysis that tries
to use the data in a more efficient way\cite{HMNT,teubner} yields
$\Delta\alpha_{had}=0.02769\pm 0.00018$. It is difficult at the
moment to beat this precision: determinations that make 
use of perturbative QCD down to lower scales in order to reduce the
error are  penalized by other uncertainties, {\it e.g.\ } on the
charm mass\cite{jeg}.
\begin{figure}[t]\center{
\psfig{figure=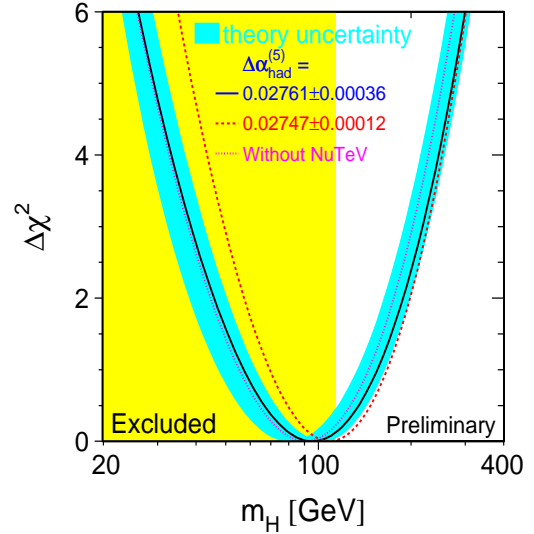,width=2.8truein,height=3.3truein}}
\caption{The parabolic darker (blue) band summarizes 
our indirect information on
  $M_H$, while the lighter shaded (yellow) area is excluded by   LEP.}
\label{fig:blue} 
\end{figure}

The  $\chi^2/$d.o.f.\ of the global fit is 25.4/15, corresponding to 
4.5\% probability. The NuTeV result shares the responsibility for the
degradation of the fit  
with another deviant measurement, that of the bottom quark
Forward-Backward asymmetry, $A_{FB}^b$, at LEP. The best
fit\cite{pippa} points to a fairly
light Higgs boson, with mass $\mh=96$~GeV, while the 95\% CL upper
bound on $\mh$, including an estimate of theoretical uncertainty,
 is about 220~GeV. As the uncertainty used for NuTeV is the one given by
the experiment, let us consider  the fit performed excluding this result.
The information on the Higgs mass is almost insensitive
to the NuTeV result ($M_H^{fit}=91$~GeV, $M_H<202~$GeV at 95\%),
 but of course the quality
of the fit improves significantly,  with $\chi^2/$d.o.f.=16.8/14,
corresponding to   26.5\%. 
One would  conclude that the SM fit is quite satisfactory. 
The direct and indirect information on the Higgs mass are summarized in
Fig.~\ref{fig:blue}, where the lighter shaded (yellow) area, $M_H<114.4$~GeV, is
excluded by LEP.  
\begin{figure*}[t]
\center{
\psfig{figure=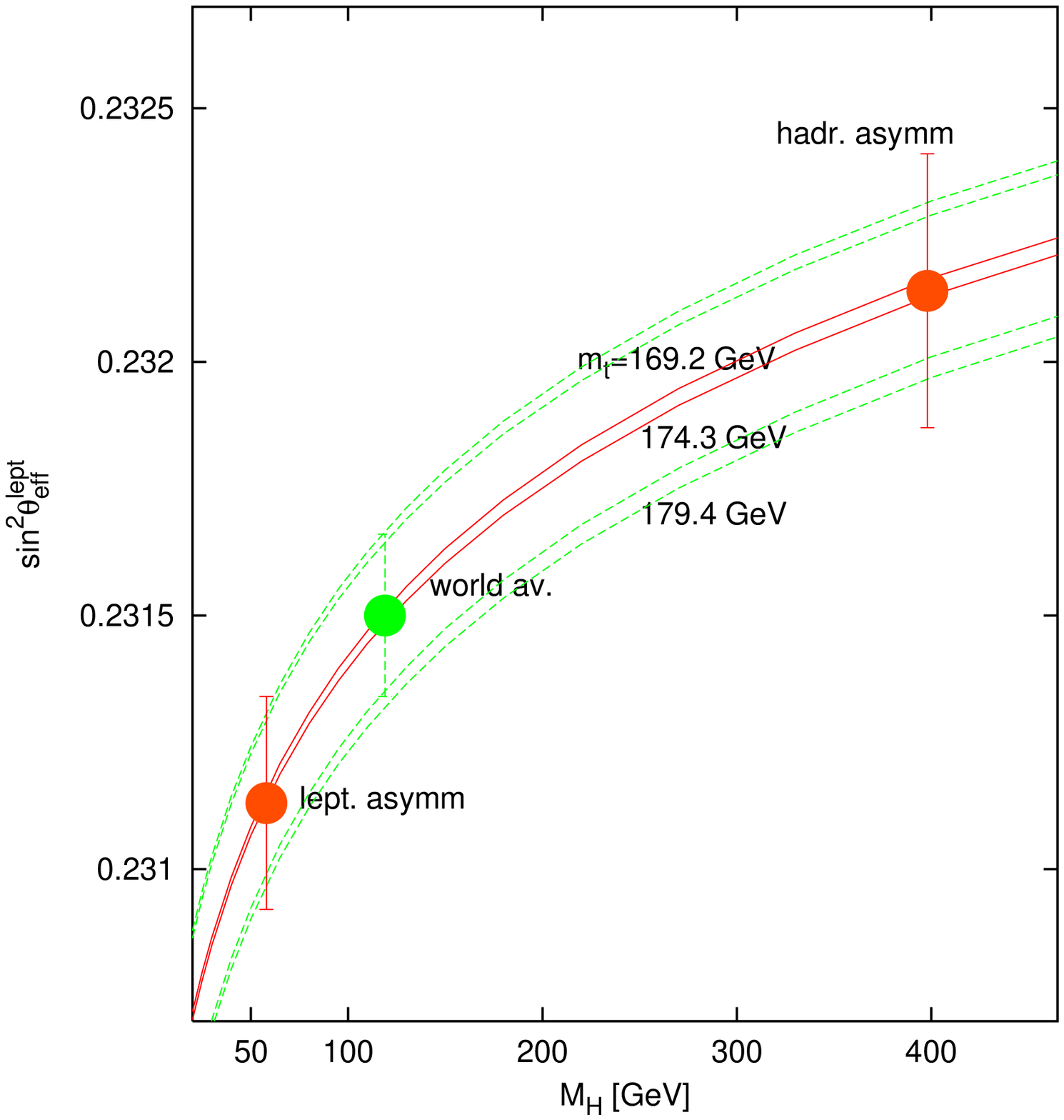,width=2.6truein,height=2.5truein}
\ \ \ \
\psfig{figure=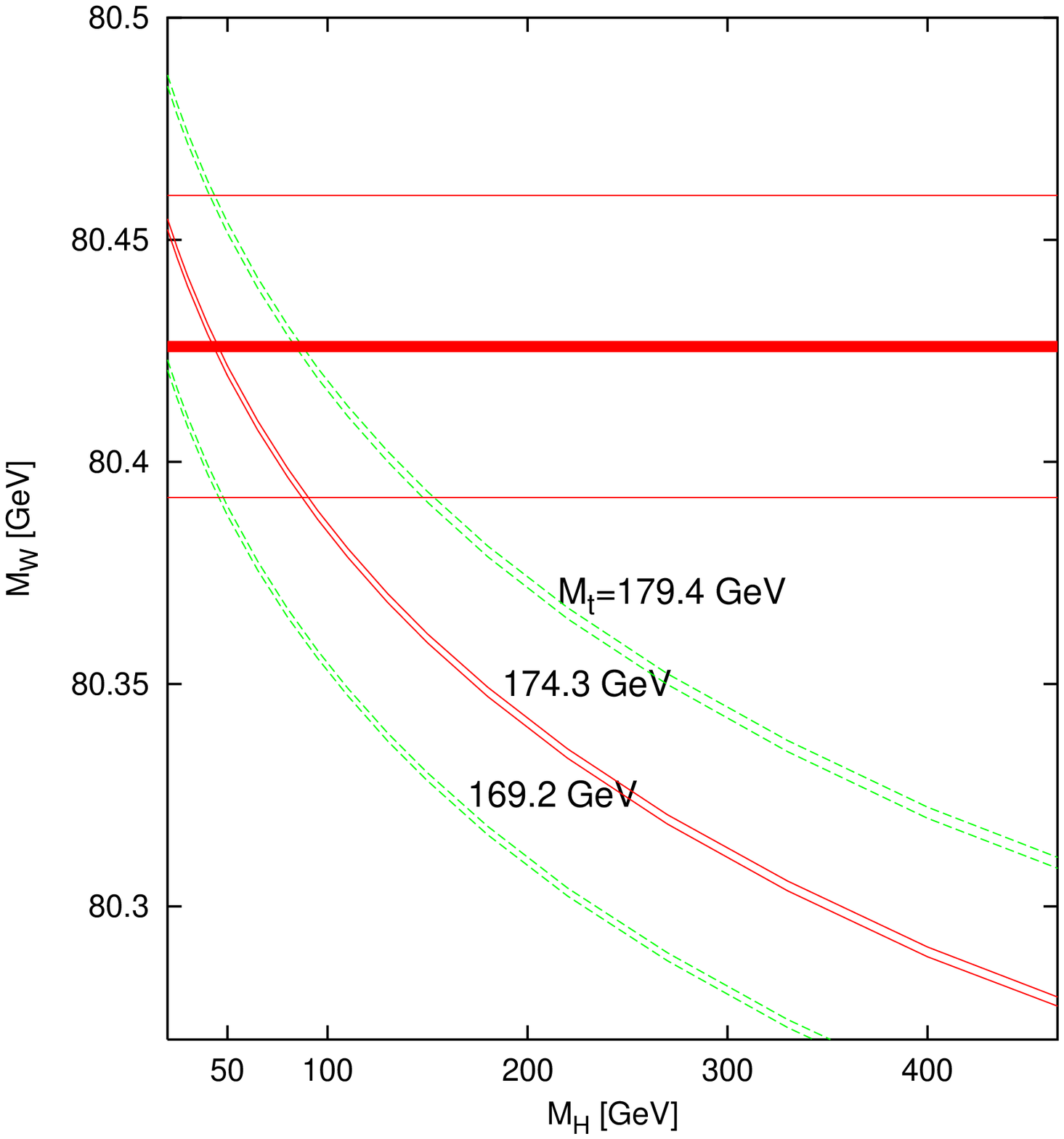,width=2.6truein,height=2.5truein}
}
\caption{ Higgs mass dependence of $\sin^2\theta_{\rm eff}^{lept}$ --
extracted from leptonic and hadronic asymmetries -- and of $M_W$,
for three $M_t$ values,  compared with the experimental values.}
\vspace{-0.4cm}
\label{fig:higgsdep} 
\end{figure*}

We have noted that, excluding NuTeV, the data are rather consistent.
But the table in Fig.~\ref{fig:pulls} contains an arbitrary set of
observables. For example, it does not include $a_\mu$, or $B\to
X_s\gamma$, which are important and  precise data. The overall conclusion would
not change, in my view.
However,  if we are interested in extracting
 information on the Higgs mass,   we should
 concentrate only on the  subset of observables that 
are really { sensitive} to $M_H$ and, because of a strong correlation, 
to the top mass, $M_t$.
Using only $M_W$, $M_t$, $\Gamma_\ell$,
the $Z$-pole asymmetries, and   $R_b$, one obtains 
$\mh^{fit}=98$~GeV,  $\mh<210 $~GeV at 95\%~CL,
and  $\chi^2/$dof=11/4, corresponding to   2.6\% probability.
In other words, the restricted fit gives the same constraints on $\mh$
 of the global fit. However, it is now obvious
 that the SM fit to the Higgs mass is {\it not}
really satisfactory.

The root of the problem is an  old $3\sigma$ discrepancy between 
the { Left-Right asymmetry, $A_{LR}$, measured by SLD
and  the Forward-Backward $b$ quark asymmetry, $A_{FB}^b$,
 measured by the  LEP experiments.
In the SM these asymmetries measure the {\it same} quantity, 
$\sin^2\theta_{\rm eff}^{lept}$, related to the lepton couplings to an
on-shell $Z^0$.
It now happens that all leptonic asymmetries, measured both at LEP and
SLD, are mutually consistent and prefer a very {\it light Higgs} mass
-- see Fig.~\ref{fig:higgsdep}. In
this sense, they are also consistent with $M_W$ measured at LEP and Tevatron.
Only the asymmetries into hadronic final states
prefer a {\it heavy Higgs} (see Fig.~\ref{fig:higgsdep}).

\begin{figure}[t]
\vspace{-1cm}
\center{
\psfig{figure=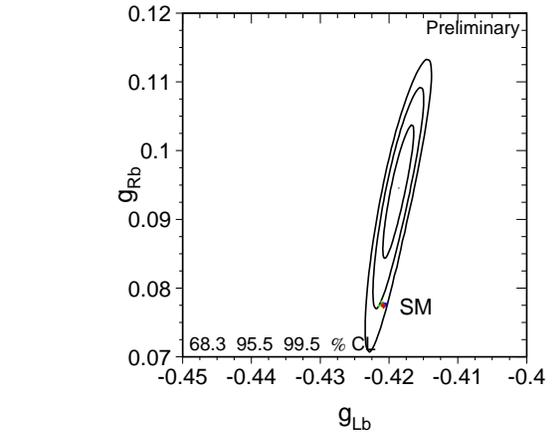,width=2.4truein}}
\caption{Fit to the left- and right-handed $b$ quark neutral current
  couplings. }
\vspace{-0.4cm}
\label{fig:glbgrb} 
\end{figure}
Since the hadronic asymmetries
are dominated by $A_{FB}^b$, and the third generation is naturally
singled out in many extensions of the SM, 
could this be  a signal of new physics in the  $b$ couplings?
After all, QCD and experimental systematics in $A_{FB}^b$ have been carefully
considered\rlap.\,\cite{pippa}
New physics in the  $b$ couplings seems unlikely for several reasons:
(i) fixing  $\sin^2\theta_{\rm eff}^{lept}$ 
at the value measured by the leptonic asymmetries, 
$A_{FB}^b$ corresponds to  a measurement of a combination of 
$b$ couplings, ${\cal{A}}_b(A_{FB}^b)=0.886\pm 0.017$; the same combination
is also tested by  $A_{LR}^{FB}$ at SLD, yielding  ${\cal{A}}_b(A_{LR}^{FB})=
0.922\pm 0.020$. 
One should compare these two values to the very precise SM prediction,
${\cal{A}}_b^{SM}=0.935\pm 0.002$: the SLD result is compatible with the SM and at
1.4\,$\sigma$ from the value extracted from $A_{FB}^b$; 
(ii)  the value  of ${\cal{A}}_b$ extracted from $A_{FB}^b$
  would require a $\sim 25\%$ 
correction to the $b$ vertex, i.e.\  tree level physics; and
(iii) $R_b$ agrees well with the SM and tests an orthogonal
combination of  $b$ couplings; it follows that  new physics should
predominantly affect the right-handed $b$ coupling, 
$|\delta g_R^b|\gg|\delta g_L^b|$, see Fig.~\ref{fig:glbgrb}. 
All this places strong restrictions on the extensions of the SM 
 that can explain 
$A_{FB}^b$.  Exotic scenarios that shift only the $b_R$ coupling include 
 mirror vector-like fermions  mixing with
the $b$  quark\rlap,\,\cite{wagner} and 
LR models that single out the third generation\rlap,\,\cite{valencia}
but even these {\it ad hoc} models have problems in 
passing all experimental tests.

We have seen that their preference for a heavy Higgs
really singles out the hadronic asymmetries.
This brings us to what can be called the {\it Chanowitz 
argument}:\cite{chanowitz,altarelli}
there are two possibilities, both involving new physics:
\begin{itemize}
\item [(a)]  $A_{FB}^b$ points to new physics; or
\item [(b)] $A_{FB}^b$ is a fluctuation or is due to unknown systematics.
\end{itemize}
In the second case it is interesting to see what happens if one excludes
the hadronic asymmetries from the above restricted Higgs mass fit. 
Not surprisingly, a  consistent picture emerges: a very light Higgs with  
{ $\mh^{fit}=42$ GeV} fits perfectly all data and one obtains an upper
bound  $\mh<120 $~GeV} at 95\%~CL. 
This would suggest new physics because the  direct lower bound on
the Higgs boson in the SM is  $\mh> 114$~GeV\rlap.\,\cite{chanowitz,altarelli} 
\begin{figure}[t]
\center{\psfig{figure=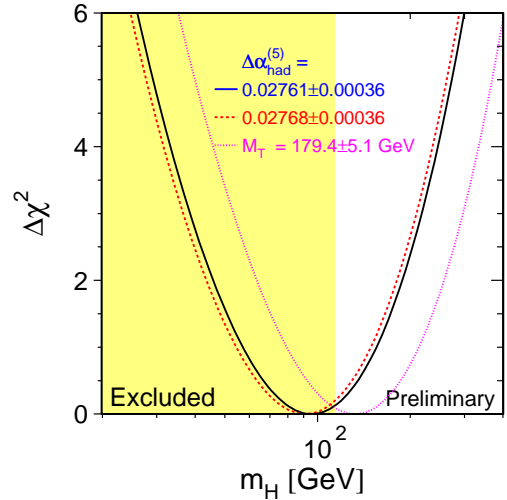,width=2.8truein}}
\caption{Effect of a 1$\sigma$ change in $M_t$ on the Higgs mass constraints. }
\vspace{-0.4cm}
\label{fig:top} 
\end{figure}

Although it may be the ringing bell for something more spectacular, 
the inconsistency with the direct lower bound is statistically rather
weak at the moment. It also marginally depends on
the value of the hadronic
contributions to $\alpha(M_Z)$ used in the fit, 
although we are already employing the most unfavorable estimate.
Similarly, current estimates of the theoretical error agree that 
it cannot shift up $\mh^{95\%}$ more than  $\sim 20$~GeV\rlap.\,\cite{freitas}
The inconsistency would be alleviated 
if the top mass turned
out to be heavier than the present central value, 
a possibility suggested by the latest D0 analysis of Run-I data
(yielding $M_t=180.1\pm 5.4$~GeV\cite{azzi}) and 
soon to be tested at the Tevatron.
Figure~\ref{fig:top} illustrates this point by showing the result of a
global fit with $M_t=179.4\pm 5.1$~GeV.

We have seen that excluding $A_{FB}^b$ (and NuTeV) from the  fit 
the quality of the fit
improves considerably, but $\mh^{fit}$ becomes very small.
Finding new physics that simulates a very light Higgs is 
much easier than fixing the two anomalies.
An example are
oblique corrections: in general it just requires $S<0~(T>0)$ or 
$\epsilon_{2,3}<0$\rlap.\,\cite{chanowitz,altarelli}
A non-degenerate unmixed fourth generation with a heavy neutrino with
$m_N\approx 50$~GeV would easily work.
More interestingly, the MSSM offers rapid decoupling (small corrections), 
  $\mw$ always higher than in the SM, and 
 $\sin^2\theta_{\rm eff}^{lept}$ lower than in the SM.
A plausible MSSM scenario involves light sneutrinos and sleptons,
 heavy squarks, and $\tan\beta\gsim 5$\rlap.\,\cite{altarelli}
\begin{figure}[t]
\center{
\psfig{figure=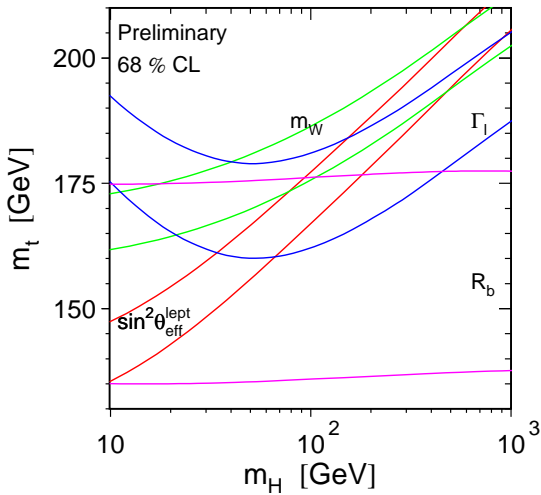,width=2.8truein}}
\caption{Constraints in the $M_t,M_H$ plane from different
  observables.}
\vspace{-0.4cm}
\label{fig:top2} 
\end{figure}

As illustrated in Fig.~\ref{fig:top}, 
the Higgs indirect determination depends strongly on
the top mass: a shift
of $+5$~GeV in $M_t$ would imply $M_H<280$~GeV instead of 200~GeV. 
A factor 2
improvement in the determination of $\alpha(M_Z)$ would lower the
95\% CL upper bound on $M_H$ by only about 5~GeV. A factor 2 
improvement in the measurement of $M_t$ would lower the
95\% CL upper bound on $M_H$ by about 35~GeV. Figure~\ref{fig:top2}
is also instructive: all the main precision observables define almost
parallel bands in the $M_t,M_H$ plane. The only important piece of
information that can, in the near future, significantly improve the
Higgs mass constraints is the top mass. A better $M_t$ measurement
would also help clarify the fate of the {\it Chanowitz argument}.

In the future, interesting new data will come from the Tevatron
($M_t$ and $M_W$), from E158 and QWeak, and later from the LHC 
and possibly from a Linear Collider. Running the latter on the
$Z^0$ peak (the {\it Giga-Z} option) would reach a new frontier in
precision physics. We will be able to exploit this precision only 
with a major effort on the theoretical side. After  years of
studies and despite
some progress\rlap,\,\cite{2loop} automatic
two-loop calculations in
the electroweak sector are still confined to special cases:
 the complete two-loop
calculation of the relation between $M_W$, $M_Z$ and $G_\mu$ has just
been completed\rlap,\,\cite{deltar} and the analogous
calculation for $\sin^2\theta_{lept}^{\rm eff}$ is nowhere in sight.

\section{{Conclusions}}
The SM works fine, but there are several areas of tension
in the data. None of them gives a convincing indication of new physics. Though each
of them could, depending on the evolution of data and theory. 

For what concerns the tests of  charged current universality, an
odd discrepancy  persists between the measurements of the Cabibbo angle 
from $K_{\ell 3}$ and nuclear $\beta$ decays. 
The situation, possibly due to 
underestimated theoretical uncertainties,  should soon be clarified
by a number of upcoming measurements.

A new global analysis of PDF's 
favors a positive strange quark asymmetry in the nucleon, that 
would reduce  the NuTeV anomaly. 
 This effect and  isospin violation in the PDF's
add a substantial uncertainty to the NuTeV result. 
Given our present understanding of the nucleon structure, the 
Paschos-Wolfenstein relation is probably  not a good 
place for electroweak precision 
physics: NuTeV may end up teaching us  more about hadronic structure than 
short-distance physics.

Revised CMD-2 data have reduced to $\approx 2\sigma$
the discrepancy between the experimental 
result for $(g-2)_\mu$ and the SM prediction based on $e^+e^-$ data. 
KLOE has given the first results with the method of radiative return, 
confirming within errors CMD-2. On the other hand, the spectral function 
extracted from $\tau$ decays still deviates significantly from $e^+e^-$ data
in a small $\sqrt{s}$ window, a rather odd result that needs to be confirmed
 and understood, probably in terms of isospin breaking.

Although the SM fit shows a clear preference for a light Higgs boson,
what we know of the Higgs mass and of the kind of new physics
we might expect depends heavily on conflicting experimental
data. Removing  the most deviant result from the SM fit leads
to a mild inconsistency with the direct lower bound on $M_H$. 
The {\it top priority} here is
a precise measurement of the top mass, and we all expect interesting
results from the Tevatron soon.

\balance

\section*{Acknowledgments}
I am grateful to A.~Ferroglia,  S.~Forte,  M.~Gr\"une\-wald,
 G.~Isidori, S.~Kretzer, K.~McFarland,
B.~Pietrzyk, A.~Polosa, G.~Rodrigo, A.~Strumia, 
T.~Teubner, Wu-Ki Tung and P.~Wells for useful discussions and communications.
This work is supported by a Marie Curie Fellowship, contract
No. HPMF-CT-2000-01048.

\end{document}